# Determination of the Low Temperature Nuclear and Magnetic Structures of $La_2O_2Se_2 \cdot Fe_2O$

David G. Free, John S. O. Evans*

Department of Chemistry, Durham University, Durham, UK, DH1 3LE

*E-mail: john.evans@durham.ac.uk

This paper describes the low temperature nuclear and magnetic structures of $La_2O_2Se_2 \cdot Fe_2O$ by analysis of X-ray and neutron diffraction data. The material has been demonstrated to order antiferromagnetically at low temperatures, with $T_N$ = ~90 K a propagation vector of $\mathbf{k}$ = (½0½), resulting in a spin arrangement similar to that in FeTe, despite there being no apparent lowering in symmetry of the nuclear structure.

## Introduction

There has been significant recent interest in mixed anion materials due largely to the discovery of superconductivity in layered oxypnictide systems.[1-3] Work on oxychalcogenides has uncovered materials with interesting magnetic and optical electronic properties, such as LaOCuS, which has been demonstrated to be a transparent *p*-type semiconductor, emitting blue light on excitation at room temperature.[4-5] These mixed anion materials often crystallise with layered structures, allowing separation of the oxide and chalcogenide / pnictide ions. A review of the structures and properties of different layered oxychalcogenide and oxypnictide materials has been given by Clarke *et al.*.[6]

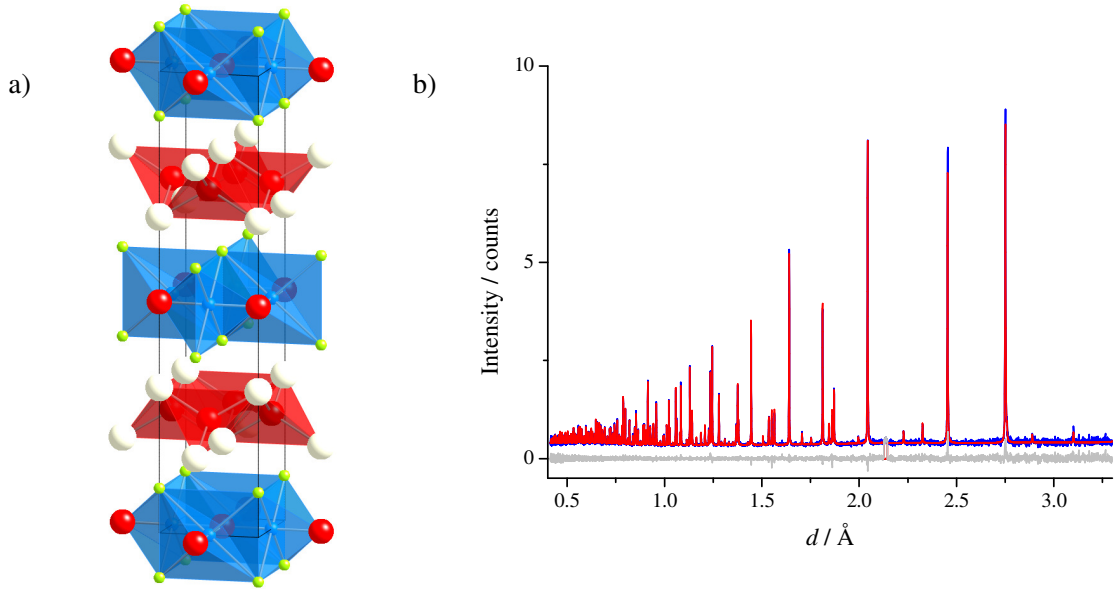

**Figure 1** (Colour online) (a) Nuclear structure of $La_2O_2Se_2.Fe_2O$ and (b) Rietveld refinement of the nuclear model of $La_2O_2Se_2.Fe_2O$ at 300 K; La = white, O = red, Se = green, Fe = blue. Data shown in (b) are from the HRPD backscattering bank (the observed pattern is in blue, calculated in red and the difference in grey)

In this paper we describe the low temperature structural and magnetic properties of the mixed anion material, $La_2O_2Se_2.Fe_2O$, Figure 1a.[7] This material, like LaOCuS and the LaOFeAs superconductors, contains layers of edge-sharing $La_4O$ tetrahedra. These are separated from $[Fe_2O]^{2+}$ layers by $Se^{2-}$ ions, which complete a square antiprismatic coordination of $La^{3+}$. The $[Fe_2O]^{2+}$ transition metal layers are a rare example of the anti-$CuO_2$ type and can also be described as a network of face-sharing octahedra, where the transition metal centred octahedron is made up of two axial oxide ions and four equatorial selenide ions. This material, and its oxysulfide analogue, have semiconducting properties and have been described as Mott insulators.[8] A large temperature independent contribution to their magnetic susceptibility with a broad maximum around 100 K, suggests antiferromagnetic ordering at low temperature. Other [$M_2$O] containing materials include Na$_2$$Pn_2$.Ti$_2$O and a recently reported family where the Na$^+$ ions have been replaced by [$A_2$O$_2$]$^{2+}$ fluorite layers.[9-11] These materials exhibit anomalous transitions in magnetic susceptibility and electrical resistivity, corresponding to CDW/SDW instabilities; magnetic ordering has not been observed from neutron powder diffraction. Also structurally related to these materials is Na$_{1.9}$Cu$_2$Se$_2$.Cu$_2$O, which contains layers of edge-sharing Cu$_4$Se tetrahedra separated from square-planar Cu$_2$O layers by Na$^+$ ions.[12]

To the best of our knowledge low temperature neutron diffraction experiments to study magnetic order in La$_2$O$_2$Se$_2$.Fe$_2$O have not been reported, however, theoretical studies of the material have suggested two possible structures, denoted AFM1 and AFM6 (see later, Figure 6), depending on the magnitude of $U$ (where $U$ is the Mott-Hubbard interaction energy).[8] Studies on the $B_2$F$_2$$Q_2$.Fe$_2$O ($B$ = Sr, Ba; $Q$ = S, Se) family, which also contains square-planar [Fe$_2$O]$^{2+}$ layers, suggest magnetic ordering occurring at 84 (neutron data) and 95-97 K (susceptibility data) for Ba$_2$F$_2$Se$_2$.Fe$_2$O and Sr$_2$F$_2$Se$_2$.Fe$_2$O, respectively.[13] Neutron data suggest an incommensurate structure for Ba$_2$F$_2$Se$_2$.Fe$_2$O, though no detailed magnetic structures were published. Calculations, again, showed the AFM1 and AFM6 arrangements to be the most stable.

**Experimental**

La$_2$O$_2$Se$_2$.Fe$_2$O for this study was prepared from stoichiometric amounts of La$_2$O$_3$ (Sigma-Aldrich, 99.9%), Fe (Aldrich, 99.9+%) and Se (Alfa-Aesar, 99.999%). The resulting powder was pressed into a 5 mm pellet and placed inside a 7 mm high density alumina crucible. This was sealed in a quartz ampoule under vacuum and heated in a furnace with the following routine: ramp to 600°C at 1°.min$^{-1}$ and dwell for 12 h, ramp to 800°C at 0.5°.min$^{-1}$ and dwell for 1 h, ramp to 1000°C at 1°.min$^{-1}$ and dwell for 12 h. After this the furnace was allowed to cool to room temperature. Analysis of the product by powder X-ray diffraction confirmed that the correct phase had been obtained. This routine was slightly different to that employed by Mayer *et al.*, as single crystals of the material were not required.[7]

Neutron data were collected using HRPD at ISIS over a time of flight window of 10-210 ms ($d$ = 0.2-16.4 Å) from 12-300K, with the sample mounted in a 5 mm vanadium slab can, for a total of 66 μAh.[14] X-ray data were collected over the sample temperature range using a Bruker d8 Advance diffractometer, with a LynxEye silicon strip detector, from 5-120° 2θ with a step size of 0.021° and collection time of 8 s per step; sample temperature was controlled using an Oxford Cryosystems PheniX CCR cryostat.[15] Neutron data were analysed over data ranges of 15-200 ms for each of three neutron banks, and 20-120° 2θ for the X-ray data. Combined X-ray and neutron refinements were performed in GSAS for the 300 and 12 K data collections.[16-17] A total of 84 variables were refined for the 300 K data (2 cell parameters, 2 atom coordinates, 5 isotropic thermal displacement parameters, 45 background parameters (XRD = 12; ND 168° = 12, 90° = 12, 30° = 9), 9 terms for TOF *x*-axis calibration (3 per ND bank), 3 absorption correction terms (1 per ND bank), a zero error term (XRD), 4 scale factors (1 per data set), 13 profile coefficients (4 XRD, 3 per ND bank). A total of 85 variables were used for the 12 K data (the additional parameter arising from description of the $Fe^{2+}$ moment). Variable temperature data shown were collected at HRPD with 6 K intervals for 2.5 μAh, and analysed using two techniques, the SEQGSAS routine in GSAS, and in TOPAS Academic using the local program *multitopas*.[16-19]

**Table 1**  Results from combined X-ray / neutron Rietveld refinements of La$_2$O$_2$Se$_2$.Fe$_2$O at 300 and 12 K, with single crystal values from the literature for comparison

|  | Mayer *et al.*[7] | *T* = 300 K | *T* = 12 K |
|---|---|---|---|
| Space group | *I*4/*mmm* | *I*4/*mmm* | *I*4/*mmm** |
| *a* / Å | 4.0788(2) | 4.084466(9) | 4.075725(6) |
| *c* / Å | 18.648(2) | 18.59798(7) | 18.53719(5) |
| *V* / Å$^3$ | 310.24 | 310.268(2) | 307.931(1) |
| La *z* / *c* | 0.18445(5) | 0.18438(3) | 0.18407(2) |
| Se *z* / *c* | 0.09669(9) | 0.09624(3) | 0.09618(2) |
| La $U_{iso}$ / 100×Å$^2$ | 0.68(4) | 0.456(8) | 0.052(4) |
| Fe $U_{iso}$ / 100×Å$^2$ | 1.96(15) | 0.785(8) | 0.186(4) |
| Se $U_{iso}$ / 100×Å$^2$ | 1.06(6) | 0.575(9) | 0.090(4) |
| O(1) $U_{iso}$ / 100×Å$^2$ | 1.0(4) | 0.60(1) | 0.249(7) |
| O(2) $U_{iso}$ / 100×Å$^2$ | 1.3(8) | 1.31(3) | 0.33(1) |
| Fe $M_x$ / $\mu_B$ | - | - | 2.82(3) |
| R$_{wp}$ / % | - | 4.96 | 3.95 |
| $\chi^2$ | - | 1.118 | 1.903 |

* The magnetic contribution to the data was modelled as a separate phase with the magnitude and direction of the moments constrained using the AFM3 model (see later).

**Table 2** Bond lengths and angles for $La_2O_2Se_2.Fe_2O$ at 12 K

| Inter-atomic distances / Å | | Bond angles / ° | |
|---|---|---|---|
| $d_{Fe-Fe}$ | 2.88198(1) | Fe-Se-Fe (1) | 64.298(7) |
| $d_{Fe-O}$ | 2.03786(1) | Fe-Se-Fe (2) | 97.62(1) |
| $d_{Fe-Se}$ | 2.7080(3) | Se-Fe-Se | 82.38(1) |
| $d_{La-O}$ | 2.3764(2) | La-O-La (1) | 105.346(7) |
| $d_{La-Se}$ | 3.3102(3) | La-O-La (2) | 118.08(2) |

## Results and Discussion

The structure was confirmed at room temperature by combined Rietveld refinement (Figure 1b) of X-ray and neutron data, and results are given in Table 1.[16-17, 20] Refinement of occupancies on individual sites, using a fixed $La^{3+}$ occupancy, confirms the expected composition of the material at 300 K (site occupancies refining to $La_2O_{1.970(4)}Se_{1.980(4)}.Fe_{1.990(2)}O_{0.999(4)}$). Refinement results were similar to those reported from single crystal studies by Mayer, although our results suggest a large thermal displacement of O(2), relative to other atoms.[7] Refinements using anisotropic thermal displacement parameters suggest that this is due to a large $U_{33}$ contribution, perpendicular to the $[Fe_2O]^{2+}$ layers. Refinement of variable temperature neutron data (Figure 2c) show that $U_{33}$ for O(2) is systematically higher than that of other sites at all temperatures.

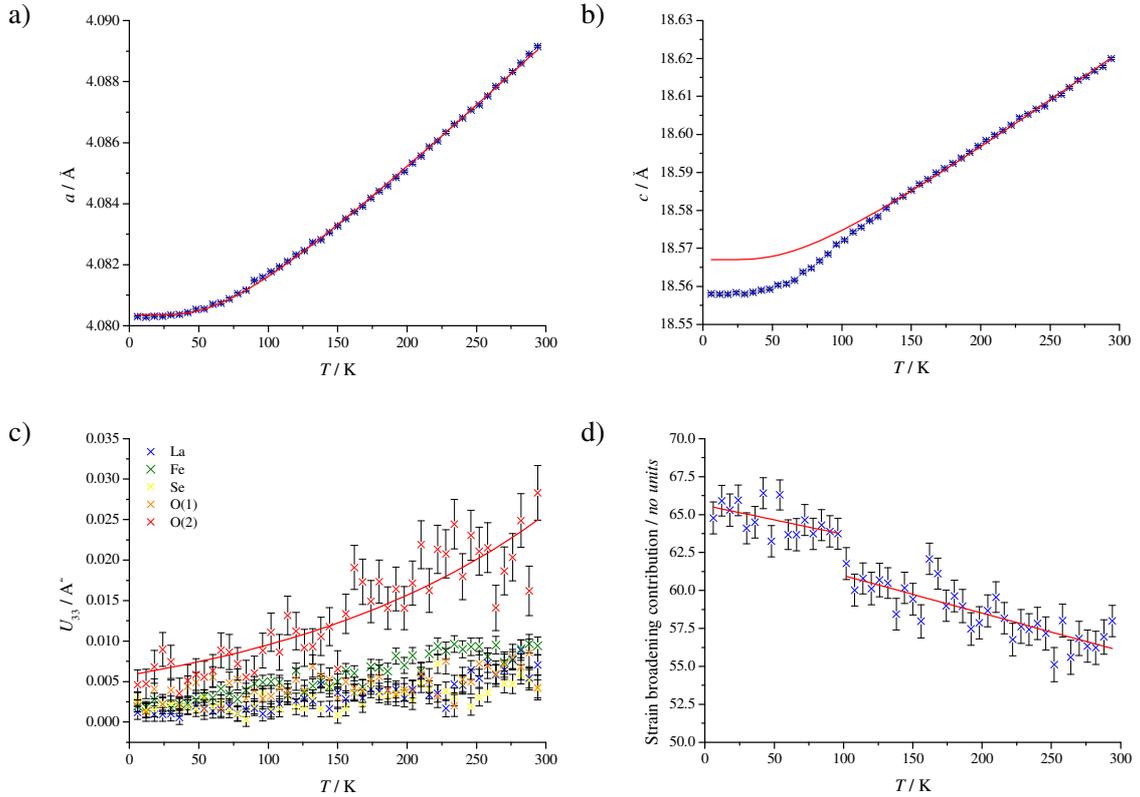

**Figure 2**  (Colour online) Plots showing the effects of temperature on the *a* (a) and *c* (b) cell parameter, $U_{33}$ values for all sites (c), and the parameter used to describe strain broadening (d); solid lines in (a) and (b) are guides to the eye (see text)

On cooling the sample below 90 K, extra peaks became visible in the 90° and 30° banks of HRPD data, at *d*-spacings of 3.01, 3.27, 3.50, 3.62, 4.44, 5.48 and 6.78 Å, that are not present in the X-ray data (Figure 3a & Figure 4), consistent with magnetic ordering of the $Fe^{2+}$ ions. These peaks were indexed using a cell with dimensions $2a \times a \times 2c$, indicating a magnetic propagation vector of ***k*** = (½0½). Plots of the nuclear cell parameters are shown in Figure 2a & b. The *a* parameter shows a continuous contraction (with perhaps a hint of discontinuity at around 90 K), whilst the *c* parameter shows a marked discontinuity at ~90 K. This is emphasised in Figure 2, where both cells have been fitted

*8*

using an Einstein model of thermal expansion. Only data above 105 K were used for *c*, so the equated $\theta_E$ of 211(4) K is dominated by the *a* axis data.[21] The thermal displacement parameter of the O(2) ion is observed to decrease continuously on cooling, in particular the $U_{33}$ parameter. These results can be seen in Figure 2c, where each atom has been modelled anisotropically, and the $U_{33}$ values have been plotted as a function of temperature. The high $U_{33}$ of O(2), corresponding to displacement above and below the $[Fe_2O]^{2+}$ plane, could indicate a local distortion of Fe-O-Fe bond away from 180°.

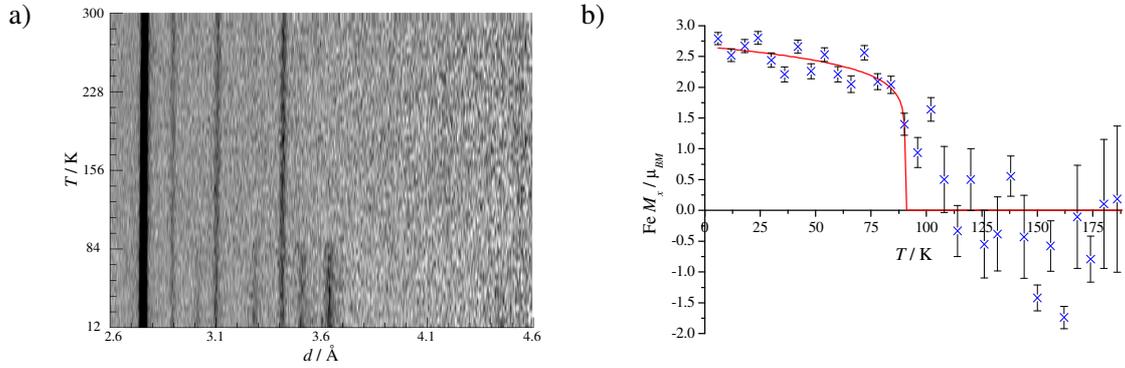

**Figure 3**  (Colour online) Plots showing the effect of temperature on (a) the evolution of the observed magnetic peaks and (b) the magnitude of the moment on the $Fe^{2+}$ ion (b); the pseudo film plot shows the 2.6-4.6 Å (90-160 ms) region of the 90° HRPD bank.

In the structurally similar pnictide superconductors, as well as in FeSe and FeTe, structural transitions from tetragonal to orthorhombic or monoclinic symmetry occur at or around $T_N$.[1, 22-28]  Despite the use of high resolution neutron (HRPD instrumental resolution $\Delta d/d \sim 4\times10^{-4}$ Å) and X-ray data, we see no evidence of peak splitting at low temperature. We note, however, that there is weak evidence of a discontinuity in the parameter used to describe peak shape strain broadening at $T_N$, which could indicate a subtle lowering of symmetry. The magnitude of the effect is, however, small and corresponds to a change in peak full width half maximum of the (200) reflection of ~4%. There is also weak evidence from the enhanced $U_{33}$ of O(2) that locally the Fe-O-Fe bond angle may deviate from 180°, which could indicate lower symmetry, though no evidence for long range order is apparent in the neutron data and $U_{33}$ varies smoothly within the quality of data available.

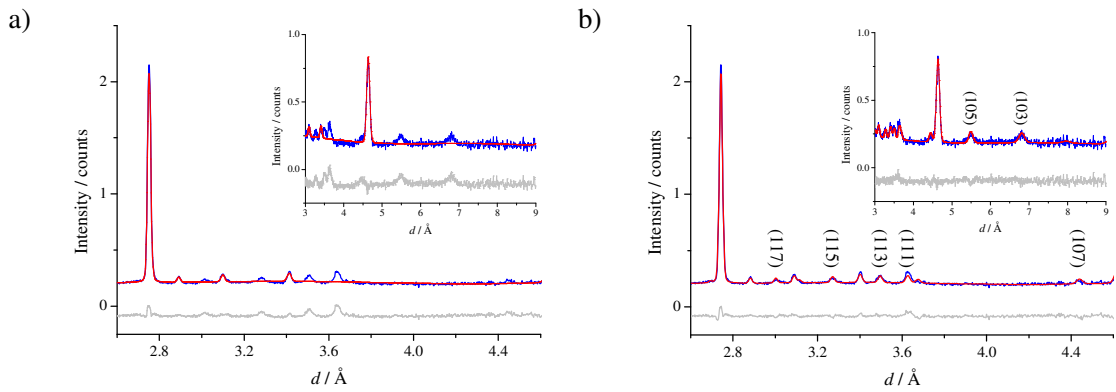

**Figure 4** (Colour online) Rietveld refinement of (a) the nuclear only and (b) nuclear plus magnetic models of $La_2O_2Se_2.Fe_2O$ at 12 K; the data shown are the 2.6-4.6 Å region from the 90° HRPD bank, and inset are data from the 3.0-9.0 Å region from the 30° bank. The observed pattern is shown in blue, calculated in red and the difference in grey; magnetic reflections are labelled.

Possible magnetic structures were investigated using a Monte-Carlo approach as implemented in SARAh Refine, which interfaces with the GSAS software suite.[16-17, 29] For $I4/mmm$ parent symmetry, with a propagation vector of $k = (½0½)$, two independent metal sites are produced (Fe(1) and Fe(2)). Results from this analysis showed good agreement with experimental data can be achieved using basis vector $\Psi_1$ associated with irreducible representation $\Gamma_2$ for Fe(1) and $\Psi_2$ associated with $\Gamma_3$ for Fe(2);[1] a contour plot of $\chi^2$ as a function of these basis vectors can be seen in Figure 5, where lower $\chi^2$ values are shown in blue and higher shown in red. Full refinement of the low temperature model using a combination of powder X-ray and neutron diffraction data, with equated moments on Fe sites, gave a good fit to the experimental data. These results are included in Table 1. The refined moment of 2.83(3) $\mu_B$ compares to values of 2.25(8) $\mu_B$ observed for Fe$_{1.068}$Te at 67 K, 0.36(5) $\mu_B$ for LaOFeAs at 8 K, and 3.32 $\mu_B$ for FeO at 77 K.[1, 27, 30] Canting of the moments was also investigated and whilst results revealed a small contribution of the moment along the $z$ axis (0.23(5) $\mu_B$), the refinement statistics were not significantly improved from the original model. Higher quality data would be needed to clarify this. The spin arrangement within a layer is shown in Figure 6c, though we note that the data of Figure 5 show that the alignment of the Fe(2) spins within a layer, relative to those of Fe(1), is not determined. The commensurate spin ordering is in contrast to that reported for Ba$_2$F$_2$Se$_2$.Fe$_2$O.[13] The observation of an antiferromagnetic arrangement of moments supports Mayer's earlier magnetic susceptibility measurements.[7]

---

[1] Fe(1)$\Gamma_2\Psi_1$ and Fe(2)$\Gamma_3\Psi_2$ are ($x$00) for $Mx,y,z$

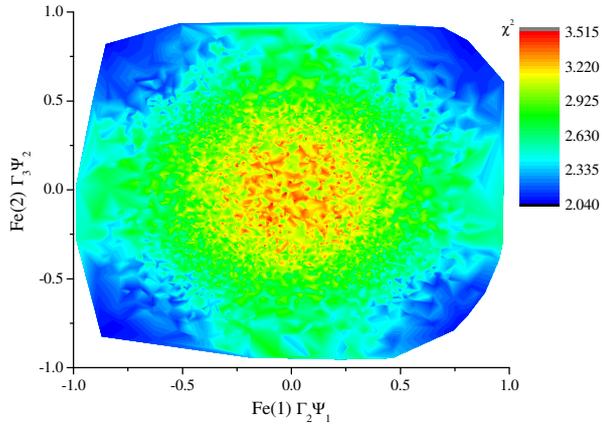

**Figure 5** (Colour online) Contour plot of the dependance of $\chi^2$ ($z$) on the $\Gamma_2\Psi_1$ and $\Gamma_3\Psi_2$ basis vectors ($x,y$), obtained by mixing of all allowed basis vectors

Kabbour *et al.* have discussed possible ordering patterns within [Fe$_2$O]$^{2+}$ layers for $A_2$F$_2$Se$_2$.Fe$_2$O ($A$ = Ba, Sr) based on DFT calculations; Zhu *et al.* have performed similar calculations on La$_2$O$_2$Se$_2$.Fe$_2$O.[8, 13] There are three important exchange interactions to consider, and we adopt the same labels as Zhu *et al.*.[2] $J_2$' is a 180° Fe-O-Fe interaction between corner-sharing octahedra, $J_2$ is the edge-sharing ~98° (see Table 2) interaction Fe-Se-Fe, and $J_1$ is a face-sharing interaction comprising Fe-Se-Fe (~64°), Fe-O-Fe (90°) and potentially direct Fe-Fe exchange; though Fe-Fe distances are ~6% larger than in FeSe$_{1-x}$ and Fe$_{1+x}$Te.[25-28] GGA+U density functional theory calculations predict $J_2$' as antiferromagnetic, $J_2$ as ferromagnetic and $J_1$ as antiferromagnetic. Predictions for $J_2$' and

---

[2] Note that different conventions are used in Kabbour's work, with $J_1$, $J_2$ and $J_2$' becoming $J_3$, $J_2$ and $J_1$, respectively.

$J_2$ are in line with simple predictions from Goodenough's rules.[31] Zhu *et al.* calculate $J_2$' = -3.28, $J_2$ = +0.78 and $J_1$ = -3.38 for $La_2O_2Se_2.Fe_2O$ for $U$ = 4.5 eV (negative $J$'s correspond to AF interactions), and predict the magnetic phase diagram for different exchange interaction strengths. For different values of $U$ the ground state is predicted to be either the AFM1 or AFM6 models of Figure 6 (Kabbour's 2F3A and 1A2F models, respectively). In the former case the $J_2$' Fe-O-Fe interactions are frustrated but $J_2$ and $J_1$ are satisfied. In the latter case half of the $J_1$ interactions are frustrated.

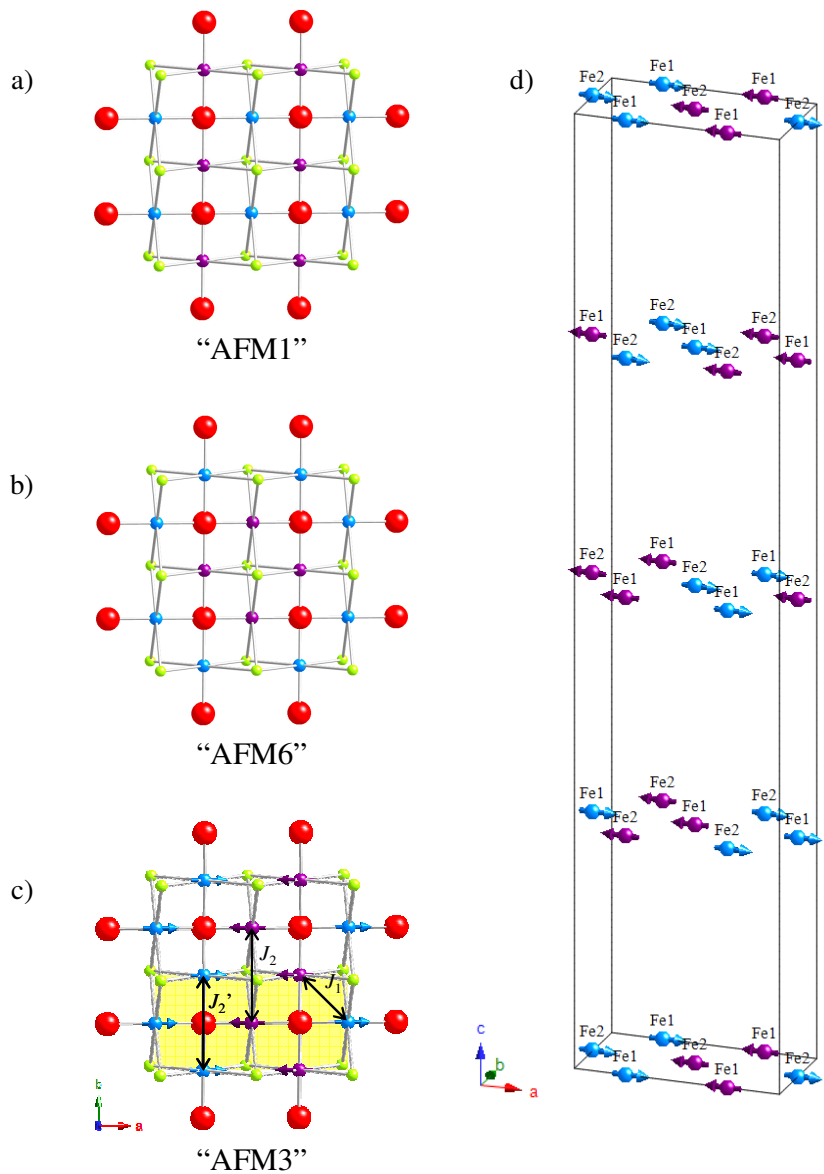

**Figure 6** (Colour online) Proposed (AFM1 (a) & AFM6 (b)) and observed (AFM3 (c) & (d)) magnetic structures of $La_2O_2Se_2 \cdot Fe_2O$; O = red, Se = green, Fe = blue/purple; the blue/purple colour of the $Fe^{2+}$ ions relates similarly orientated moments; the magnetic cell in (c) is shaded

The observed magnetic structure doesn't correspond to either of the predicted structures and is shown in Figure 6 (c & d). Spins are aligned ferromagnetically along the *b* direction and antiferromagnetically along *a*. Note that the relative orientation of spins on the Fe(1) and Fe(2) lattices within a layer is not determined. Along the doubled cell edge Fe-O-Fe interactions are antiferromagnetic as expected, whereas they are ferromagnetic along *b*. Conversely $J_2$ interactions are satisfied along the short axis, *b*, but not along *a*. Half of the $J_1$ interactions are similarly frustrated, as seen in the LaOFeAs systems. The magnetic structure is similar to that observed for $Fe_{1+x}Te$ samples, with $x < 0.1$ (Figure 7a & b), which have the same **k** = (½0½) propagation vector and spin arrangement, with the majority spin direction in the *ab* plane, but predominantly aligned along the short axis.[27-28, 32-33] In this system magnetic ordering is accompanied by a clear phase transition to a monoclinic $P2_1/m$ cell with $a/b = 1.01$ and $\beta = 89.2°$, and hence the observed magnetic ordering can be described by a single irreducible representation. The symmetry lowering leads to exchange interactions differing in different directions, as labelled in Figure 7b, and the region of the magnetic phase diagram in which the AFM3 structure is stable has been discussed by Fang *et al.*[33]

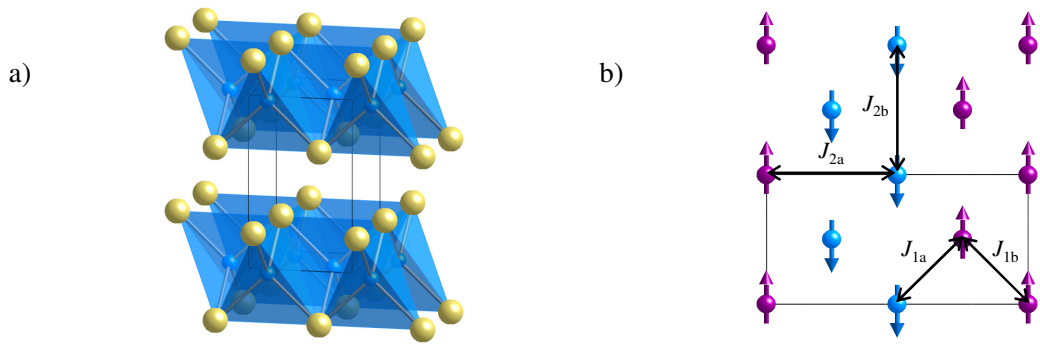

**Figure 7**  (Colour online) Nuclear (a) and magnetic (b) structures of and $Fe_{1.086}Te$; Fe = blue/purple, Te = gold

## Conclusions

In conclusion, neutron diffraction data have shown that the magnetic moments within $La_2O_2Se_2.Fe_2O$ order with an AFM3 arrangement at ~90 K, similar to that in $Fe_{1.086}Te$. This arrangement has all interactions frustrated and is contrary to calculations that predicted the AFM1 and AFM6 models to be the most likely arrangements.[8, 13] We see no evidence of peak splitting at $T_N$ which might reflect a lowering of symmetry, though there is a marked change in the *c*/*a* ratio at this temperature. The observation of antiferromagnetism at this temperature is in agreement with susceptibility measurements published by Mayer *et al*.

## Acknowledgements

We wish to thank Dr. A. Daoud-Aladine and Dr. E. E. McCabe (Durham) for measurements on the HRPD at the ISIS facility, and also the EPSRC for funding.

**References**

1. C. de la Cruz, Q. Huang, J. W. Lynn, J. Li, W. Ratcliff II, J. L. Zarestky, H. A. Mook, G. F. Chen, J. L. Luo, N. L. Wang and P. Dai, Nature **453**, 899-902 (2008).

2. Y. Kamihara, H. Hiramatsu, M. Hirano, K. Kawamura, H. Yanagi, T. Kamiya and H. Hosono, J. Am. Chem. Soc. **128**, 10012-10013 (2006).

3. H.-H. Wen, G. Mu, L. Fang, H. Yang and X. Zhu, Europhys. Lett. **82**, 17009 (2008).

4. K. Ishikawa, S. Kinoshita, Y. Suzuki, S. Matsuura, T. Nakanishi, M. Aizawa and Y. Suzuki, J. Electrochem. Soc. **138** (4), 1166 (1991).

5. K. Ueda, S. Inoue, S. Hirose, H. Kawazoe and H. Hosono, Appl. Phys. Lett. **77** (17), 2701 (2000).

6. S. J. Clarke, P. Adamson, S. J. C. Herkelrath, O. J. Rutt, D. R. Parker, M. J. Pitcher and C. F. Smura, Inorg. Chem. **47**, 8473-8486 (2008).

7. J. M. Mayer, L. F. Schneemeyer, T. Siegrist, J. V. Waszczak and B. v. Dover, Angew. Chem. Int. Ed. Engl. **31** (12), 1645-1647 (1992).

8. J. X. Zhu, R. Yu, H. Wang, L. L. Zhao, M. D. Jones, J. Dai, E. Abrahams, E. Morosan, M. Fang and Q. Si, arXiv:0912.4741v1 (2009).

9. R. H. Liu, Y. A. Song, Q. J. Li, J. J. Ying, Y. J. Yan, Y. He and X. H. Chen, Chem. Mater. **22**, 1503-1508 (2010).

10. T. C. Ozawa and S. M. Kauzlarich, Chem. Mater. **13**, 1804-1810 (2001).

11. T. C. Ozawa, R. Pantoja, E. A. Axtell, S. M. Kauzlarich, J. E. Greedan, M. Bieringer and J. W. Richardson, J. Solid State Chem. **153** (2), 275-281 (2000).

12. Y. B. Park, D. C. Degroot, J. L. Schindler, C. R. Kannewurf and M. G. Kanatzidis, Chem. Mater. **5** (1), 8-10 (1993).

13. H. Kabbour, E. Janod, B. Corraze, M. Danot, C. Lee, M.-H. Whangbo and L. Cario, J. Am. Chem. Soc. **130**, 8261-8270 (2008).

14. R. Ibberson, W. I. F. David and K. S. Knight, *The High Resolution Neutron Powder Diffractometer (HRPD) at ISIS - A User Guide*. (ISIS Crystallography, Didcot, 1992).

15. Oxford Cryosystems, (2007).

16. A. C. Larson and R. B. Von Dreele, (Los Alamos National Laboratory: Los Alamos, 2004).

17. B. H. Toby, in *J. Appl. Cryst.* (2001), Vol. 34, pp. 210-221.

18. J. S. O. Evans, (University of Durham, 1999).

19. A. A. Coelho, (Bruker AXS, Karlsruhe, 2007).

20. H. M. Rietveld, J. Appl. Cryst. **2**, 65 (1969).

21. K. Wang and R. R. Reeber, Appl. Phys. Lett. **76** (16), 2203-2204 (2000).

22. J. Zhao, Q. Huang, C. de la Cruz, S. Li, J. W. Lynn, Y. Chen, M. A. Green, G. F. Chen, G. Li, Z. Li, J. L. Luo, N. L. Wang and P. Dai, Nature Mater. **7**, 953-959 (2008).

23. Q. Huang, J. Zhao, J. W. Lynn, G. F. Chen, J. L. Luo, N. L. Wang and P. Dai, Phys. Rev. B **78**, 054529 (2008).

24. J. Zhao, Q. Huang, C. de la Cruz, J. W. Lynn, M. D. Lumsden, Z. A. Ren, J. Yang, X. Shen, X. Dong, Z. Zhao and P. Dai, Phys. Rev. B **78**, 132504 (2008).

25. S. Margadonna, Y. Takabayashi, M. T. McDonald, K. Kasperkiewicz, Y. Mizuguchi, Y. Takano, A. N. Fitch, E. Suard and K. Prassides, Chem. Comm., 5607-5609 (2008).

26. F. C. Hsu, J. Y. Luo, K. W. Yeh, T. K. Chen, T. W. Huang, P. M. Wu, Y. C. Lee, Y. L. Huang, Y. Y. Chu, D. C. Yan and M. K. Wu, P. Natl. Acad. Sci. USA **105** (38), 14262-14264 (2008).

27. S. Li, C. de la Cruz, F. Q. Huang, Y. Chen, J. W. Lynn, J. Hu, Y. L. Huang, F. C. Hsu, K. W. Yeh, M. K. Wu and P. Dai, Phys. Rev. B **79**, 054503 (2009).

28. D. Fruchart, P. Convert, P. Wolfers, R. Madar, J. P. Senateur and R. Fruchart, Mater. Res. Bull. **10** (3), 169-174 (1975).

29. A. S. Wills, Physica B **276-278**, 680-681 (2008).

30. W. L. Roth, Phys. Rev. **110**, 1333 (1958).

31. J. B. Goodenough, *Magnetism and the Chemical Bond*, 1st ed. (John Wiley & Sons, New York - London, 1963).

32. W. Bao, Y. Qui, Q. Huang, M. A. Green, P. Zajdel, M. R. Fitzsimmons, M. Zhernenkov, S. Chang, M. Fang, B. Qian, E. K. Vehstedt, J. Yang, H. M. Pham, L. Spinu and Z. Q. Mao, Phys. Rev. Lett. **102**, 241001 (2009).

33. C. Fang, B. A. Bernevig and J. Hu, Europhys. Lett. **86**, 67005 (2009).